\documentclass[prd,twocolumn,superscriptaddress,altaffilletter,showpacs]{revtex4}
\usepackage{amssymb,amsmath}
\usepackage{comment}
\usepackage{graphicx}
\usepackage{csquotes}
\usepackage{tikz}
\usetikzlibrary{snakes}
\usepackage{lipsum, babel}
\DeclareUnicodeCharacter{2212}{\textendash}

\begin{document}

\title{Mass, charge, and distance to Reissner-Nordstr\"{o}m black hole \\in terms of directly measurable quantities}
\author{Gerardo Morales-Herrera}
\email{gerardo.moraleshe@alumno.buap.mx}
\affiliation{Facultad de Ciencias F\'{i}sico Matem\'{a}ticas, Benem\'{e}rita Universidad Aut\'{o}noma de Puebla, Ciudad Universitaria, Puebla, CP 72570, Puebla, Mexico}
\author{Pablo Ortega-Ruiz}
\email{pablo.ortegar@alumno.buap.mx}
\affiliation{Facultad de Ciencias F\'{i}sico Matem\'{a}ticas, Benem\'{e}rita Universidad Aut\'{o}noma de Puebla, Ciudad Universitaria, Puebla, CP 72570, Puebla, Mexico}
\author{Mehrab Momennia}
\email{mmomennia@ifuap.buap.mx, momennia1988@gmail.com}
\affiliation{Instituto de F\'{\i}sica, Benem\'erita Universidad Aut\'onoma de Puebla,\\
Apartado Postal J-48, 72570, Puebla, Puebla, Mexico}
\affiliation{Instituto de F\'{\i}sica y Matem\'{a}ticas, Universidad Michoacana de San
Nicol\'as de Hidalgo,\\
Edificio C--3, Ciudad Universitaria, CP 58040, Morelia, Michoac\'{a}n, Mexico}
\author{Alfredo Herrera-Aguilar}
\email{aherrera@ifuap.buap.mx}
\affiliation{Instituto de F\'{\i}sica, Benem\'erita Universidad Aut\'onoma de Puebla,\\
Apartado Postal J-48, 72570, Puebla, Puebla, Mexico}

\date{\today }

\begin{abstract}

In this paper, we employ a general relativistic formalism and develop new theoretical tools that allow us to analytically express the mass and electric charge of the Reissner-Nordstr\"{o}m black hole as well as its distance to a distant observer in terms of few directly observable quantities, such as the total frequency shift, aperture angle of the telescope, and redshift rapidity. Our analytic and concise formulas are valid on the midline, and the redshift rapidity is a relativistic invariant observable that represents the evolution of the frequency shift with respect to the proper time in the Reissner-Nordstr\"{o}m spacetime. This procedure is applicable for particles undergoing circular motion around a spherically symmetric and electrically charged black hole, which is the case for accretion disks orbiting supermassive black holes hosted at the core of active galactic nuclei. Although this type of black hole is expected to be electrically neutral, our results provide a novel method to measure the electric charge of the Reissner-Nordstr\"{o}m black hole, hence can shed some light on this claim. Besides, these results allow us to measure the mass of the black hole and its distance from the Earth, and we can employ the general formulas in black hole parameter estimation studies.

\vskip3mm

\end{abstract}

\maketitle

%: circular equatorial orbits}

%%%%%%%%%%%%%%%%%%%%%%%%%%%%%%%%%%%%%%%%%%%%%%%%%%%%%%%%%%%%%%%%%%%%
\section{Introduction}
In recent years, black hole physics has become a very active research field, and robust theoretical predictions \cite{penrose} and observational evidence from gravitational waves \cite{GWmergBH} as well as images of black holes' shadows \cite{Akiyama,EventH3} indicate these remarkable supercompact bodies, predicted by Einstein's General Relativity theory, do exist in the cosmos. 

Black holes are simple objects, completely described by only a handful of parameters, namely mass $M$, electric charge $Q$, and angular momentum $J$, known as the No-hair theorem \cite{nohair1,nohair3, nopelo}. A black hole that contains all three parameters is called a Kerr-Newman black hole that reduces to a charged static Reissner-Nordstr\"{o}m (hereafter RN) black hole in the spin-zero limit $J=0$. 

Although the electric charge in static RN black holes is often considered to be negligible \cite{RN}, there is no conclusive evidence to assert that these bodies are not present in nature (see \cite{ExtremeRN} for observational signatures of the RN black hole), hence electrically charged black holes have been extensively studied in the literature so far . For instance, based on gravitational retro-lensing images, a method has been proposed to determine the electric charge of a RN black hole \cite{C1}, and to do so, the authors suggested employing the shadow size of the black hole. In addition, different methods for constraining the electric charge of the galactic center black hole Sgr A* are presented in \cite{C2,C3,C4,EHTimageConstraint}. Despite presenting considerably good approximations for the black hole mass, there is no reliable value for the electric charge. However, it was claimed that a Reissner-Nordström black hole with a significant electric charge provides a better fit to the observational data for the black hole shadow in comparison with the Schwarzschild black hole \cite{C5}. In the present study, we propose a general relativistic method to determine the electric charge of an RN black hole and its mass in terms of directly measurable quantities.

Lately, the study of the observational frequency shift and its relation to black hole parameters and cosmological constant has been investigated in \cite{d2,b, c, d, e, f},  in contrast to the previous attempts based on the kinematic redshift that is not an observable quantity \cite{Lopez, Becerril, Becerril2, AHerrera}. A proper attempt to comprehend the importance of the frequency shift can be achieved by studying the motion of test massive particles and light in order to understand the physical properties of the region hosting them.

In \cite{b}, the authors have developed a new general relativistic method to analytically express the mass and spin of Kerr black holes in terms of the total observational frequency shift experienced by the photons emitted from massive geodesic particles as well as the radius of the emitters' orbits (for equatorial circular motion). The main contribution of this novel method is to decode the information about the relativistic effects present in the black hole environment which is hidden in the total frequency shift experienced by the out-coming photons. This innovative method aims to extend the physical interpretation of the frequency shift to General Relativity and improve the Newtonian picture of gravitation which is valid just when the test particle is orbiting far away from the central source. 

More recently, it was shown that generalizing this approach to the Kerr black hole in asymptotically de Sitter spacetime allows us to extract the Hubble constant from the total observational redshift \cite{c}, hence enabling us to measure the cosmological constant and black hole parameters, simultaneously (see also \cite{KdSshadow} for a recent additional avenue for testing the cosmological constant by using the Kerr-de Sitter black hole shadow). In addition, this general relativistic method has been applied to polymerized black holes to express the black hole parameters in terms of the total redshift \cite{d}. Furthermore, this approach has been extended to general spherically symmetric spacetimes in order to extract the information on free parameters of black hole solutions in alternative theories of gravitation \cite{e}. 

On the other hand, practically, the mass-to-distance ratio of 17 supermassive black holes hosted at the core of the active galactic nuclei (AGNs) has been estimated in \cite{towards, Villalobos, deby, Villalobos2} by employing this general relativistic approach. In these parameter estimation studies of the supermassive Schwarzchild black holes, just the mass-to-distance ratio $M/D$ of these compact objects has been estimated due to the fact that this ratio is degenerate in this general relativistic formalism. However, more recently, a method has been proposed to overcome the mass-to-distance ratio degeneracy problem in the Schwarzchild background based on the \textit{redshift rapidity} such that the mass of the Schwarzchild black hole and its distance from the Earth were expressed just in terms of directly observable elements \cite{f}. The redshift rapidity is observable and represents the evolution of the frequency shift with respect to the proper time in the black hole background.

In this paper, we employ a similar general relativistic method in order to express the RN black hole parameters, mass and electric charge as well as the distance of the black hole to the observer, as functions of directly observational and measurable quantities. We do so by introducing the redshift rapidity in the RN background which is the derivative of the redshift with respect to proper time and is a general relativistic invariant observable quantity.  Hence, with the aid of the redshift rapidity and the total frequency shifts, we extract concise and elegant analytic formulas for the parameters of the RN black hole in terms of these observables as well as the aperture angle of the telescope characterizing the emitter position. 

It is worth mentioning that our analytic formulas enable us to obtain the RN black hole mass and charge as well as its distance from the Earth in a completely general relativistic framework. In this direction, one of our main goals in developing this approach to more general black hole solutions is improving the precision in measuring cosmic distances with respect to previous post-Newtonian methods that compute the angular-diameter distance to several megamaser systems, such as UGC 3789 \cite{MCP2}, NGC 6264 \cite{MCPV}, NGC 5765b \cite{MCPVIII}, NGC 4258 \cite{Humphreys2013,Reid2019}, and CGCG 074-064 \cite{MCPXI}.

The outline of the paper is as follows. In Sec. \ref{Geodesics}, we elaborate on the geodesic motion of massive and null particles moving in the RN background. We derive expressions for the nonvanishing components of four-velocity of massive test particles and the photons' four-momentum as well as for their impact parameter in terms of RN black hole mass and charge. Next, in Sec. \ref{FrequencyShift}, we present a brief overview of our general relativistic method in the RN background and use the results of Sec. \ref{Geodesics} to obtain the frequency shifts of test particles circularly orbiting the RN black hole for an arbitrary point on the circular path. Then, we present a couple of expressions for the mass-to-distance ratio and charge-to-distance ratio of the RN black hole in terms of observational quantities on the midline.  In Sec. \ref{rrapidity}, we define the redshift rapidity as the proper time evolution of the frequency shift in the RN background.  Then, in Sec. \ref{Disentangle}, we employ the results of the redshift and the redshift rapidity obtained, respectively, in Secs. \ref{FrequencyShift} and \ref{rrapidity}, to express the black hole mass $M$, electric charge $Q$, and its distance $D$ from the Earth only in terms of observational elements. Finally, in Sec. \ref{COnclusion}, we conclude with some final remarks related to the developed general relativistic formalism as well as a brief discussion of the applications and relevance of our results.

\section{Geodesic motion in Reissner-Nordstr\"{o}m spacetime}
\label{Geodesics}

%\textbf{Here, you introduce the RN black hole spacetime. There are some formulas and citation for guide.}

In this section, we set the context of our general relativistic method and derive the nonvanishing components of the four-velocity of a test particle revolving the black hole spacetime in terms of the metric parameters by analyzing its equation of motion. Also, we obtain the components of the four-momentum for null particles and express the impact parameter in terms of RN black hole mass and charge. The four-velocity elements of the massive test particles and their impact parameter will be useful to construct an expression for the frequency shift in the RN background. 

Consider an electrically charged, static, and spherically symmetric RN black
hole with the following line element in the Schwarzschild coordinates $\left( t,r,\theta ,\varphi \right) $
[we use $c=1=G=k_c$ units] 
\begin{equation}
ds^{2}=g_{tt}dt^{2}+g_{rr}dr^{2}+g_{\theta \theta }d\theta ^{2}+g_{\varphi
\varphi }d\varphi ^{2},  \label{metric}
\end{equation}%
with the metric components%
\begin{equation}
    g_{tt}=-f(r),\quad g_{rr}=\frac{1}{f(r)},\\
    \label{gtt}
\end{equation}% 
\begin{equation}
    g_{\theta \theta }=r^2 ,\quad g_{\varphi\varphi}=r^2sin ^{2}\theta,
    \label{gphiphi}
\end{equation}
in which the metric function  $f(r)$ is given by
\begin{equation}
    f(r)= 1-\frac{2M}{r}+\frac{Q^2}{r^2},
    \label{f}
\end{equation}
where $M$ and $Q$ are the mass and the electric charge of the black hole, respectively. 

The RN spacetime has a curvature singularity at the origin $r=0$, whereas the two coordinate singularities characterized by the roots of the metric function $f(r_{\pm})=0$, specify the Cauchy horizon and the event horizon surfaces
\begin{equation}
    r_{\pm} = M\pm (M^2-Q^2)^{1/2}.
\end{equation}

The particles located in the vicinity of an RN black hole feel the curvature of spacetime produced by the black hole mass and charge. They encode information about the spacetime curvature and the RN black hole parameters in the frequency shift of emitted photons \cite{b,AHerrera}. Hence, further study regarding the motion of the massive/massless geodesic particles moving under the RN background (\ref{metric}) is necessary which we shall describe in the coming subsections.

\subsection{Geodesics of massive particles}
\label{nonnull}

A neutral massive test particle following geodesics in the RN spacetime has a four-velocity of the form
\begin{equation}
U^\mu = (U^t, U^r, U^\theta, U^\varphi), \quad U^{\mu}=\frac{dx^\mu}{d\tau},
\end{equation}
with the normalization condition 
\begin{equation}
U^\mu U_\mu = -1.
\label{normal}
\end{equation}

In this study, we are interested in the test particle's orbits to be confined in a plane with the motivation of describing thin accretion disks circularly orbiting supermassive black holes at the center of many AGNs. Conveniently, due to the spherical symmetry of the spacetime (\ref{metric}), this plane can be chosen to be the equatorial plane $\theta=\pi/2$ without loss of generality. Therefore, the polar component of the four-velocity vanishes ($U^{\theta} = 0$), and the $g_{\varphi\varphi}$-component of the metric (\ref{gphiphi}) reduces to $r^2$. The Killing vector fields $\psi^{\nu}=(0,0,0,1)$ and $\xi^{\nu}=(1,0,0,0)$ of the RN metric, associated with the rotational and temporal symmetries, allow us to relate  the azimuthal and temporal components of the four-velocity to, respectively, the angular momentum and total energy per unit rest mass of the orbiting particle as follows
\begin{equation}
    L=\frac{\Bar{L}}{m}=\psi _{\mu}U^{\mu}=g_{\mu \nu}\psi ^{\nu}U^{\mu}=g_{\varphi \varphi}U^{\varphi}
    \label{killin2},
\end{equation}
\begin{equation}
    E=\frac{\Bar{E}}{m}=-\xi _{\mu}U^{\mu}=-g_{\mu \nu}\xi ^{\nu}U^{\mu}=-g_{tt}U^t.
    \label{killn}
\end{equation}

Now, by substituting $U^{t}$ and $U^{\varphi}$ from the above-mentioned equations into the equation of motion (\ref{normal}), we get
\begin{equation}
-\frac{1}{2}g_{tt}g_{rr}(U^r)^2-\frac{g_{tt}}{2}-\frac{g_{tt}}{2g_{\varphi\varphi}}L^2=\frac{E^2}{2}.
\label{veff}
\end{equation}

The relation (\ref{veff}) takes the form of the energy conservation law for a non-relativistic particle moving under the following effective potential 
\begin{equation}
    V_{eff}=-\frac{1}{2}g_{tt} \left(1+\frac{L^2}{g_{\varphi\varphi}}\right).
\end{equation}

We notice from Eq. (\ref{veff}) that we can define 
\begin{equation}
    g_{rr}(U^r)^2=-1 -\frac{E^2}{g_{tt}}-\frac{L^2}{g_{\varphi \varphi}} \equiv V_r(r),
\label{RadPot}
\end{equation}
which is a function of $r$ only.
Now, we examine the special case of circular orbits where the radial component $U^r$ in Eq. (\ref{veff}) vanishes. As a consequence, the radial function $V_r(r)$ is zero, as for its first derivative with respect to $r$
%hola xd
\begin{equation}
    V_r(r)= -1 -\frac{E^2}{g_{tt}}-\frac{L^2}{g_{\varphi \varphi}}=0,
   \label{vefd}
\end{equation}
\begin{equation}
     V'_{r}(r)=\frac{E^2}{g_{tt}^2}\frac{\partial g_{tt}}{\partial r }+\frac{L^2}{g_{\varphi \varphi}^2}\frac{\partial g_{\varphi \varphi}}{\partial r }=0.
     \label{veffprima}
\end{equation}

Hence, by substituting the RN metric components $g_{tt}$ and $g_{\varphi \varphi}$ from Eqs. (\ref{gtt})-(\ref{gphiphi})  in the aforementioned equations, we can solve them to get the following relations for $E$ and $L$ in terms of the metric function
\begin{equation}
\left. E=f\sqrt{\frac{2}{2f-r\frac{\partial f}{\partial r}}}\right|_{r=r_e} ,
\label{energy}
\end{equation}
\begin{equation}
    \left. L =(\pm)r\sqrt{\frac{r\frac{\partial f}{\partial r}}{2f-r\frac{\partial f}{\partial r}}} \right|_{r=r_e},
    \label{energy2}
\end{equation}
where the index $e$ refers to the orbital radius of the emitter.

Now, it is convenient to notice that from Eqs. (\ref{f}), (\ref{killin2}), (\ref{killn}), (\ref{energy}), and (\ref{energy2}), we can build the temporal and azimuthal components of the four-velocity in terms of the RN black hole parameters $M$ and $Q$ as follows  
\begin{equation}\label{Ut}
     U^t_e=\frac{E}{f(r_e)}=\frac{r_e}{\sqrt{r_e^2-3Mr_e+2Q^2}}, 
\end{equation}
\begin{equation}\label{Uphi}
     U^{\varphi}_e=\frac{L}{r_e^2}=\frac{1}{r_e}\sqrt{\frac{Mr_e-Q^2}{r_e^2-3Mr_e+2Q^2}}.
\end{equation}

The relation between the metric parameters and these nonvanishing four-velocity components of the test particle is a key point for further developments in the following sections that will allow us to decode the information of the spacetime curvature caused by the mass and charge of the RN black hole.

Besides, the stability criterion is given by the second derivative of the effective potential (or equivalently the radial potential $V_r(r)$ given in Eq. (\ref{RadPot})) such that the radius of the orbiting particle must satisfy $V''_{r} \leq 0$ to ensure stability of the orbit. In the case where the second derivative of the effective potential is identically zero, the corresponding radius is the innermost stable circular orbit (hereafter ISCO).  The second-order derivative of $V_r(r)$ is given by
\begin{align}\label{estable}
    V''_{r}&= -\frac{2E^2}{g^3_{tt}}\left(\frac{\partial g_{tt}}{\partial r}\right)^2+\frac{E^2}{g^2_{tt}}\frac{\partial^2g_{tt}}{{\partial r}^2} \notag\\
    &-\frac{2L^2}{g^3_{\varphi\varphi}}\left(\frac{\partial g_{\varphi\varphi}}{\partial r}\right)^2+\frac{L^2}{g^2_{\varphi\varphi}}\frac{\partial^2g_{\varphi\varphi}}{{\partial r}^2}.
\end{align}

To determine the radius of the ISCO, we substitute $g_{tt}$, $g_{\varphi \varphi}$, $E$, and $L$ from, respectively, Eqs. (\ref{gtt}), (\ref{gphiphi}), (\ref{energy}), and (\ref{energy2}) into Eq. (\ref{estable}) and set to zero to get the following cubic equation for $r$
\begin{equation}
    Mr^3-6M^2r^2+9MQ^2r-4Q^4 =0,
\end{equation}
whose only real solution is 
\begin{equation}
    r_{ISCO}=\frac{a^2Q^{\frac{4}{3}}-3M^{2/3}Q^2+4M^{8/3}}{M^{1/3}a}+2M,
\label{Risco}
\end{equation}
where
\begin{equation}
    a=\left(\frac{8M^4}{Q^2}-9M^2+2+\sqrt{5M^4-9M^2Q^2+4Q^4} \right)^{1/3},
\end{equation}
and represents the boundary for which any inward or outward perturbation to a circular orbit will result in spiraling into the black hole or escaping from it. In this work, we only take into account test particles on circular orbits with the radii $r_e \geq r_{ISCO}$.

\subsection{Geodesics of null particles}

Here, we analyze the motion of the photons emitted by the massive particles described in the previous section. These photons move with a four-momentum $k^{\mu}=(k^t,k^r,k^{\theta},k^{\varphi})$ through null geodesics satisfying
\begin{equation}
    k^{\mu}k_{\mu}=0.
    \label{momento}
\end{equation}

Due to the spherical symmetry of the spacetime, the motion of the photons has certain conserved quantities, namely the total energy $E_{\gamma}$ and angular momentum $L_{\gamma}$, that are obtained through the following expressions
\begin{equation}
     E_{\gamma}=-\xi _{\mu}k^{\mu}=-g_{\mu \nu}\xi ^{\nu}k^{\mu}=-g_{tt}k^t, 
\end{equation}
\begin{equation}
     L_{\gamma}=\psi _{\mu}k^{\mu}=g_{\mu \nu}\psi ^{\nu}k^{\mu}=g_{\varphi \varphi}k^{\varphi}.
\end{equation}

With the help of these relations, we can express the $k^t$ and $k^\varphi$ components of the four-momentum in terms of $E_{\gamma}$ and $L_{\gamma}$
\begin{equation}
    \left. k^t=-\frac{E_\gamma}{g_{tt}}\right|_{r=r_e}=\frac{E_{\gamma}}{f(r_e)},
\label{energyphoton}
\end{equation}
\begin{equation}
     \left. k^{\varphi}=\frac{L_{\gamma}}{g_{\varphi \varphi}}\right|_{r=r_e}=\frac{L_{\gamma}}{r_e^2sin ^{2}\theta},
     \label{eneryphoton2}
\end{equation}
where the index $\gamma$ refers to the photons. Then, by expanding the expression (\ref{momento}) and substituting the values of $k^t$ (\ref{energyphoton}) and $k^\varphi$ (\ref{eneryphoton2}), we obtain

\begin{equation}
    0 = g_{rr}(k^r)^2 + \frac{E_{\gamma}^2}{g_{tt}} + \frac{L_{\gamma}^2}{g_{\varphi \varphi}},
\label{k}
\end{equation}
where we take into account that in the equatorial plane, the polar component of the four-momentum is null ($k^\theta=0$). Hence, from Eq. (\ref{k}), it is possible to notice that the square of the radial component of the four-momentum is
\begin{equation}
    \left. (k^r)^2=-\frac{g_{tt}L_{\gamma}^2+g_{\varphi\varphi}E_{\gamma}^2}{g_{rr}g_{tt}g_{\varphi\varphi}}\right|_{r=r_e}.
    \label{4momento}
\end{equation}

\begin{figure*}[t]
    \centering
    \includegraphics[scale=0.8]{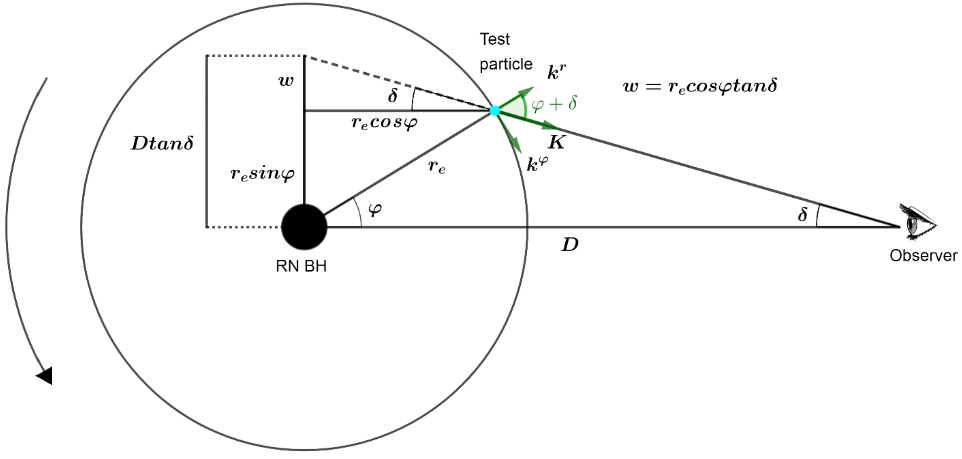}
    \caption{Illustration of the vector $K$, and geometric relations between the azimuthal angle $\varphi$ and the aperture angle $\delta$ for an observer placed in the equatorial plane at a distance $D$.}
    \label{Angulos}
\end{figure*}

Now, by considering Fig. \ref{Angulos}, we introduce the vector $K$ described by the following geometric relations
\begin{equation}
    k^r=Kcos(\varphi + \delta),
    \label{kcos}
\end{equation}
\begin{equation}\label{kr}
r_ek^{\varphi}=Ksin(\varphi+\delta),
\end{equation}
where $\varphi$ is the azimuthal angle. Therefore, from expressions (\ref{kcos}) and (\ref{kr}), the two-dimensional auxiliary vector $K$ is defined by
\begin{equation}
    K^2=(k^r)^2+r_e^2(k^{\varphi})^2,
    \label{4momento2}
\end{equation}
where by substituting the components of the four-momentum found in (\ref{eneryphoton2}) and (\ref{4momento}), we can build $K^2$ in terms of the energy and angular momentum of the photons
\begin{equation}
    K^2=-\frac{g_{tt}L_{\gamma}^2+g_{\varphi\varphi}E_{\gamma}^2}{g_{rr}g_{tt}g_{\varphi\varphi}}+r_e^2\left(\frac{L_{\gamma}}{g_{\varphi\varphi}}\right)^2.
    \label{K2}
\end{equation}

On the other hand, from Eqs. (\ref{4momento}) and (\ref{kcos}), it is straightforward to find that 
\begin{equation}
    K^2=-\frac{g_{tt}L_{\gamma}^2+g_{\varphi\varphi}E_{\gamma}^2}{g_{rr}g_{tt}g_{\varphi\varphi}cos^2(\varphi+\delta)}.
    \label{K1}
\end{equation}

Equating Eqs. (\ref{K2}) and (\ref{K1}), we obtain the following relation
\begin{equation}
 b_{\varphi}^2g_{tt}\left[g_{\varphi\varphi}tan^2(\varphi+\delta)+r_e^2g_{rr}\right]+\left[g_{\varphi\varphi}tan(\varphi+\delta)\right]^2=0,
\end{equation}
where $b_{\varphi}\equiv\frac{L_{\gamma}}{E_{\gamma}}$ is the impact parameter \cite{RefResp}. 
By solving the previous relation for $b_{\varphi}$, we obtain an expression for the bending of the light experienced by photons emitted by test particles located at any point in a circular orbit. The impact parameter in terms of the black hole parameters is given by
\begin{equation}\label{lbp}
     b_{\varphi}=-\frac{r_esin(\varphi+\delta)}{\sqrt{f(r_e)sin^2(\varphi+\delta)+cos^2(\varphi+\delta)}}.
\end{equation}

It is worth mentioning that this parameter takes into account the bending of light due to the gravitational field in the vicinity of the RN black hole produced by mass and charge, and it is preserved along the whole null geodesics followed by photons from their emission till their detection. Therefore, since both $E_{\gamma}$ and $L_{\gamma}$ are constants of motion, we have $b_{\varphi,e}=b_{\varphi,d}$, where $b_{\varphi,e}$ refers to the impact parameter when photons are emitted and $b_{\varphi,d}$ when they are detected.

Besides, as is illustrated in Fig. \ref{Angulos}, the angles $\varphi$ and $\delta$ are geometrically related by
\begin{equation}\label{rel}
D\tan{\delta}=r_e\sin{\varphi}+r_e\cos{\varphi}\tan{\delta},
\end{equation}
where after doing some straightforward modifications, this relation reduces to
\begin{equation}\label{rel2}
D\sin{\delta}=r_e\sin(\varphi+\delta).
\end{equation}

By solving the previous equation, we can express $\delta$ in terms of the rest of the parameters. Generally, equality (\ref{rel2}) has four solutions, and we choose the physical one as below
\begin{equation}\label{deltaphirel}
    \delta(\varphi)=\arccos{\left(\frac{D-r_e\cos{\varphi}}{\sqrt{D^2+r^2_e-2Dr_e\cos{\varphi}}}\right)}.
\end{equation}

\section{Frequency shift in Reissner-Nordstr\"{o}m background}

\label{FrequencyShift}

%\textbf{Here you explain the frequency shift in RN background for a general
%point on the circuar orbit. Then, present approximations on the midline.}
%%%%%%%%%%%%%%%%%%%%% z & zDot Fig %%%%%%%%%%%%%%%%%%%%%%%%
\begin{figure*}[t]
\centering
\includegraphics[scale=.9]{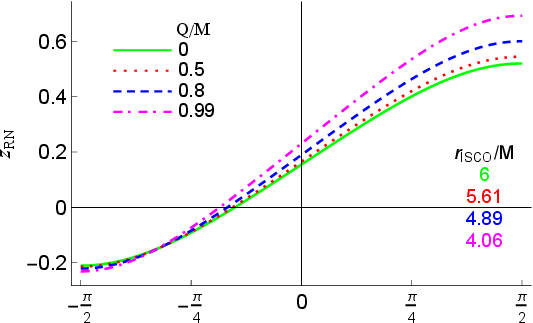}
\includegraphics[scale=.95]{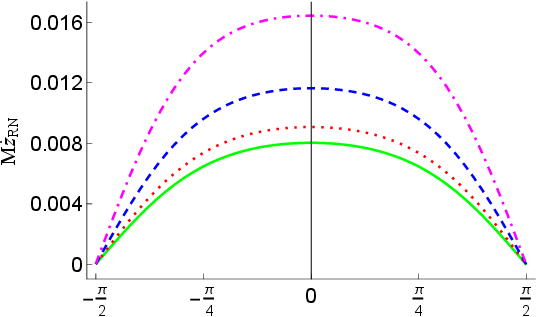}
\includegraphics[scale=.9]{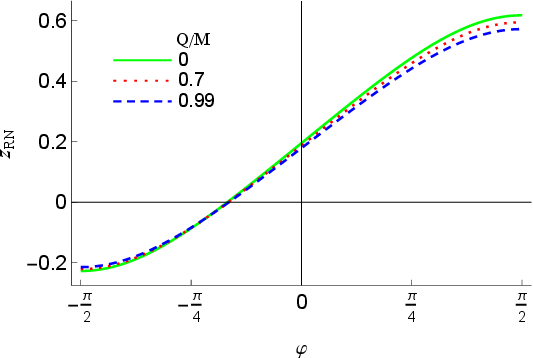}
\includegraphics[scale=.95]{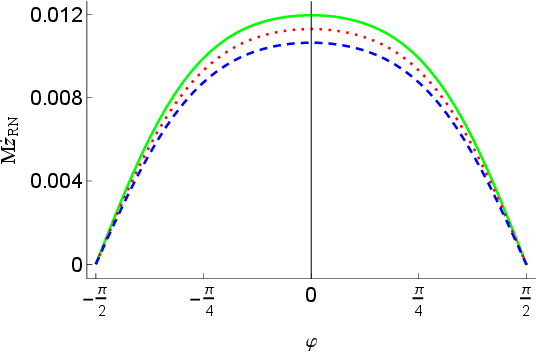}
\caption{The frequency shift $z_{RN}$ and the redshift rapidity $\dot{z}_{RN}$ versus the azimuthal angle in the RN background for $r_{e}=2r_{ISCO}$, $D=10^3r_{ISCO}$ (upper panels) and $r_{e}=10M$, $D=10^4M$ (lower panels), and different values of the electric charge $Q$. The shift in the frequency is maximal on the midline where $\varphi\approx\pm \pi /2$, whereas the maximum redshift rapidity is at the line of sight $\varphi=0$. The continuous green curves correspond to the frequency shift and the redshift rapidity in the Schwarzschild spacetime. To plot these curves, we replaced $r_{ISCO}$ (for upper panels) and $\delta$ (for all plots) given by, respectively, Eqs. (\ref{Risco}) and (\ref{deltaphirel}) into Eqs. (\ref{Redshift}) and (\ref{RRgeneral}).}
\label{zDotFig}
\end{figure*}
%%%%%%%%%%%%%%%%%%%%%%%%%%%%%%%%%%%%%%%%%%%%%%%%%%

Once the photons have been emitted, they experience the gravitational influence of the black hole, hence, from emission to detection, they would undergo a shift in their frequency. The frequency of a photon at some position $x_{p}^{\mu }=\left(
x^{t},x^{r},x^{\theta },x^{\varphi }\right) \mid _{p}$\ reads
\begin{equation}
\omega _{p}=-\left( k_{\mu }U^{\mu }\right) \mid _{p}\,,  \label{freq}
\end{equation}
where the index $p$ refers to either the point of emission $x_{e}^{\mu }$ or
detection $x_{d}^{\mu }$ of the photon.

The most general expression for shifts in the frequency $\omega _{p}$ in
static and spherically symmetric backgrounds of the form (\ref{metric}) can
be written as 
\begin{eqnarray}
1 &+&z_{_{BH}}\!=\frac{\omega _{e}}{\omega _{d}}  \notag \\
&=&\frac{(E_{\gamma }U^{t}-L_{\gamma }U^{\varphi
}-g_{rr}U^{r}k^{r}-g_{\theta \theta }U^{\theta }k^{\theta })\mid _{e}}{%
(E_{\gamma }U^{t}-L_{\gamma }U^{\varphi }-g_{rr}U^{r}k^{r}-g_{\theta \theta
}U^{\theta }k^{\theta })\mid _{d}}\,.  \label{GeneralShift}
\end{eqnarray}

In the RN background, equation (\ref{GeneralShift}) takes a specific form that depends on the components of the four-momentum of emitted photons and the four-velocity of the emitter, which were previously found in Eqs. (\ref{Ut})-(\ref{Uphi}), (\ref{energyphoton})-(\ref{eneryphoton2}), and (\ref{4momento}). However, we recall that the radial $U_p^r$ and polar $U_p^{\theta}$ components of the four-velocity vanish for both the emitter and the detector. Besides, taking into account that the detector is placed at a large distance $r_d=D>>r_e$, its four-velocity reduces to 
\begin{equation}
    U_d^{\mu} = (1, 0, 0, 0),
\end{equation}  
since $U^\varphi_d$ approaches zero and the temporal term $U^t_d$ is equal to the unity for $r_d\to\infty$, because the receptor does not experience the relativistic effects present in the vicinity of the central compact body.
Finally, by considering the restrictions mentioned above, the general expression (\ref{GeneralShift}) reduces to 
\begin{align} \label{Redshift}
    1+z_{RN} &=\frac{E_{\gamma}U^t_e-L_{\gamma}U^{\varphi}_e}{E_{\gamma}}=U^t_e-b_{\varphi}U^{\varphi}_e\notag\\
    &=\frac{1}{\sqrt{1-3\tilde{M}+2\tilde{Q}^2}}\times \notag\\
    &\left[1+\frac{\sqrt{\tilde{M}-\tilde{Q}^2}\sin(\varphi+\delta)}{\sqrt{\cos^2(\varphi+\delta)+f(r_e)\sin^2(\varphi+\delta)}}\right],
\end{align}
for the frequency shift in RN spacetime and reduces to the Schwarzschild black hole case \cite{f} in the limit $\tilde{Q}=0$, as we expected. In this relation, $\tilde{M}=M/r_e$, $\tilde{Q}=Q/r_e$, and we introduced Eqs. (\ref{Ut})-(\ref{Uphi}) and (\ref{lbp}) in the last equality. In the frequency shift formula (\ref{Redshift}), $z_{RN}$ and $\delta$ are directly observable elements, $r_e$ and $\varphi$ are unobservable quantities, and $M$ and $Q$ are the parameters of interest that should be obtained.

The left panels of Fig. \ref{zDotFig} illustrate the redshift formula (\ref{Redshift}) versus the azimuthal angle $\varphi$ for different values of the electric charge $Q$. This figure shows the changes of frequency shift in the RN spacetime with the motion of the massive test particle. As one can see, the frequency shift either increases or decreases with increasing $Q$, depending on the radius of the emitter. In other words, if the radius of the emitter is fixed at $r_e=2r_{ISCO}$ for each black hole, the frequency shift increases as the electric charge increases (because the radius of ISCO decreases). However, if we fix the radius of the emitter for every black hole at $r_e=10M$, increasing $Q$ leads to decrement in the redshift. Besides, on the midline where $\varphi\approx\pm \pi /2$, the total frequency shift is maximal, hence easier to be identified observationally.

In addition, the expression (\ref{Redshift}) represents the frequency shift experienced by the out-coming photons emitted by the test particles moving around an RN black hole with circular orbits, and we can identify their gravitational contribution and their kinematic contribution to the total redshift, which are as follows
\begin{equation}
    z_g = U_e^t - 1,
\end{equation}
\begin{equation}
    z_{kin}=-b_{\varphi}U_e^{\varphi}.
\end{equation}

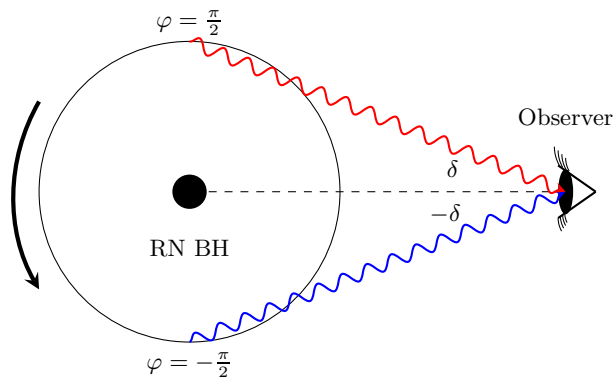
\begin{figure*}
\begin{center}
\begin{tikzpicture}
    \draw[black, dashed, thin](10,0)--(5,0);
    \draw[black, thin] (5,0) circle (2);
    \filldraw[color=black, fill=black, ultra thick] (5,0) circle (0.2);
    \fill[black] (10,0) ellipse (0.1 and 0.3);
    \draw[black, thick] (10.4,0)--(9.9,0.35);
    \draw[black, thick] (10.4,0)--(9.9,-0.35);
    \draw[snake=snake,blue, ->,>= stealth, thick] (5,-2)--(10,0);
    \draw[snake=snake,red, ->,>= stealth, thick] (5,2)--(10,0);
    \draw [thin] (9.85,0.6) arc [start angle=180, end angle=210, radius=0.5cm];
    \draw [thin] (9.9,0.55) arc [start angle=180, end angle=210, radius=0.45cm];
    \draw [thin] (9.95,0.5) arc [start angle=180, end angle=210, radius=0.4cm];
    \draw [thin] (10,0.45) arc [start angle=180, end angle=210, radius=0.35cm];
    \draw [thin] (9.9,-0.35) arc [start angle=180, end angle=210, radius=0.35cm];%eyelashes down
    \draw [thin] (9.93,-0.32) arc [start angle=180, end angle=210, radius=0.33cm];
    \draw [thin] (9.96,-0.3) arc [start angle=180, end angle=210, radius=0.3cm];
    \draw [thin] (9.99,-0.3) arc [start angle=180, end angle=210, radius=0.22cm];
    \node at (10,1) {Observer};
    \node at (5,-0.75) {RN BH};
    \node at (5,2.25) {$\varphi=\frac{\pi}{2}$};
    \node at (5,-2.3) {$\varphi=-\frac{\pi}{2}$};
    \node at (8.5,0.3) {$\delta$};
    \node at (8.4,-0.3) {$-\delta$};
    \draw [ultra thick, -stealth] (3,1.2) arc [start angle=150, end angle=210, radius=2.5cm];
\end{tikzpicture}
\end{center}
 \caption{Pictorial representation of the frequency shifts of two different light rays emitted by timelike sources orbiting an RN black hole in an equatorial circular geodesic motion when the total redshift and blueshift reach their maxima at $\varphi=\pm \pi/2$. An observer, which is also located in the equatorial plane, detects these photons coming from the midline.}
 \label{fig:maxshifts}
\end{figure*}

Now, in order to be able to express the mass-to-distance ratio and the charge-to-distance ratio in terms of observables, let us take into account the values of the redshift and blueshift when the orbiting particles are placed on the midline where $\varphi=\pm \pi /2$, as shown in Fig. \ref{fig:maxshifts}. At these points, the impact parameter is maximized and Eq. (\ref{Redshift}) takes the following form
\begin{align}\label{blueandred}
    1+z_{RN_{1,2}} &=\frac{1}{\sqrt{1-3\tilde{M}+2\tilde{Q}^2}}\times \notag\\
    &\Biggl[1\pm\frac{\sqrt{\tilde{M}-\tilde{Q}^2}\cos{\delta}}{\sqrt{\sin^2{\delta}+(1-\tilde{M}+\tilde{Q}^2)\cos^2{\delta}}}\Biggr],
\end{align}
%\begin{figure*}
    %\centering
    %\includegraphics[scale=0.231]{FIg 2.png}
    %\caption{Test particles on the midline rotating counterclockwise.}
    %\label{fig:2}
%\end{figure*}
where the plus sign corresponds to the receding (redshifted) particle denoted by index $1$ and the minus sign refers to the approaching (blueshifted) particle denoted by index $2$.

From Eq. (\ref{blueandred}), it is possible to set a system of equations as follows
\begin{equation}\label{rb1}
 RB=\frac{1}{1-3\tilde{M}+2\tilde{Q}^2}\left[\frac{\sec^2{\delta}-3\tilde{M}+2\tilde{Q}^2}{\sec^2{\delta}-2\tilde{M}+\tilde{Q}^2} \right],
\end{equation}
\begin{equation}\label{rb2}
    (R+B)^2=\frac{4}{1-3\tilde{M}+2\tilde{Q}^2},
\end{equation}
where we defined $R=1+z_{RN_1}$ and $B=1+{z_{RN_2}}$. Hence, we can solve these equations to obtain $\tilde{M}$ and $\tilde{Q}$ in terms of observables as below
\begin{equation}
    \tilde{M}=\frac{4}{(R+B)^2}-\Sigma+2\sec^2{\delta}-1,
\label{tildeM}    
\end{equation}
\begin{equation}
    \tilde{Q}^2 = \frac{8}{(R+B)^2}-\frac{3}{2}\Sigma+3\sec^2{\delta}-2,
    \label{tildeQ}
\end{equation}
where
\begin{equation}
    \Sigma = \frac{4+(R+B)^2\tan^2{\delta}}{2RB}.
\end{equation}

We recall that we consider the distance $D$ between the black hole center and the detector to be very large, thus $\delta$ is small (see Fig. \ref{Angulos}). Therefore, the relation (\ref{rel2}) leads to the following approximation of the emitter radius
\begin{equation}
    r_e\approx D\delta\left(1+\frac{\delta^2}{3}\right),
    \label{rdelta}
\end{equation}
for $\delta\ll 1$ on the midline. Then,
we introduce the approximation (\ref{rdelta}) into Eqs. (\ref{tildeM})-(\ref{tildeQ}) and perform Taylor expansion about a small delta up to the fourth order to get
\begin{align}\label{m}
    \frac{M}{D}\approx&\left(1-\frac{2}{RB}+\frac{4}{(R+B)^2}\right)\delta\\
    &+\left(\frac{1}{3}-\frac{2}{3RB}+\frac{4}{3(R+B)^2}-\frac{(R-B)^2}{2RB}\right)\delta^3\notag,
\end{align}
\begin{align}\label{q}
    \frac{Q^2}{D^2}\approx&\left(1-\frac{3}{RB}+\frac{8}{(R+B)^2}\right)\delta^2\\
    &+\left(\frac{2}{3}-\frac{2}{RB}+\frac{16}{3(R+B)^2}-\frac{3(R-B)^2}{4RB}\notag
    \right)\delta^4,
\end{align}
where denote the mass-to-distance ratio (\ref{m}) and the charge-to-distance ratio (\ref{q}) which are only 
functions of the directly observable elements $\{R,B,\delta\}$ that should be measured 
on the midline.

\section{Redshift rapidity in Reissner-Nordstr\"{o}m background}
\label{rrapidity}

In this section, we develop further the formalism in order to express the mass $M$, charge $Q$, and distance $D$ to the RN black hole only in terms of observational quantities by employing the redshift rapidity, a new concept recently introduced in the Schwarzschild black hole spacetime \cite{f}.

Hence, we define the redshift rapidity as the proper time evolution of the frequency shift $z_{RN}$ (\ref{Redshift}) in the RN background as follows 
\begin{equation}\label{redrapi}
      \dot{z}_{RN,e}=\frac{dz_{RN}}{d\tau}=\frac{d}{d\tau}\left(U^t_e-b_\gamma U^\varphi_e\right),
\end{equation}
which is the redshift rapidity at the point of emission.

However, the redshift rapidity needs to be measured from the Earth. Therefore, we use the chain rule in order to rewrite Eq. (\ref{redrapi}) at the observer position as \cite{f}
\begin{equation}\label{redrapit}
    \dot{z}_{RN}=\frac{dz_{RN}}{dt}=\frac{d\tau}{dt}\frac{dz_{RN}}{d\tau}=\frac{1}{U^t_e}\frac{d}{d\tau}\left(U^t_e-b_\gamma U^\varphi_e\right),
\end{equation}
which is an observable quantity that we measure here on the Earth. Hence, for massive geodesic particles circularly orbiting the black hole in the equatorial plane, the redshift rapidity (\ref{redrapit}) reduces to
\begin{equation}
    \dot{z}_{RN}=-\frac{db_\gamma}{d\tau}\frac{U^\varphi_e}{U^t_e},
\end{equation}
since the temporal component $U^t_e$ (\ref{Ut}) and azimuthal component $U^\varphi_e$ (\ref{Uphi}) of the four-velocity are constant quantities for circular orbits whereas the impact parameter (\ref{lbp}) depends on time through $\delta$ and $\varphi$. By making use of the chain rule, we have \cite{f} 
\begin{equation}\label{totredrapi}
    \dot{z}_{RN}=-\left(\frac{\partial b_\gamma}{\partial\varphi}+\frac{\partial b_\gamma}{\partial\delta}\frac{\partial\delta}{\partial\varphi}\right)\frac{(U^\varphi_e)^2}{U^t_e},
\end{equation}
where we used $U^\varphi_e=\frac{d\varphi}{d\tau}\Big\rvert_{r=r_e}$. Hence, by performing the derivatives of the impact parameter (\ref{lbp}) and the $\delta(\varphi)$-relation (\ref{deltaphirel}) presented in (\ref{totredrapi}), the redshift rapidity for an arbitrary point on the circular orbit reads 
\begin{align}\label{RRgeneral}
    \dot{z}_{RN}=&\frac{D}{r_e}\left[\frac{1-\tilde{M}-f(r_e)}{\sqrt{2f(r_e)+\tilde{M}-1}}\right]\left[ \frac{D-r_e\cos{\varphi}}{r_e^2+D^2-2r_eD\cos{\varphi}} \right]\times\notag \\
    &\frac{cos(\varphi + \delta)}{\left[ f(r_e)sin^2(\varphi+\delta)+cos^2(\varphi+\delta)\right]^{3/2}},
\end{align}
where reduces to the redshift rapidity in the Schwarzschild background \cite{f} for $\tilde{Q}=0$, as it should be.

The right panels of Fig. \ref{zDotFig} show the behavior of the redshift rapidity versus the azimuthal angle $\varphi$ for various values of the electric charge $Q$. Similar to the frequency shift case, the redshift rapidity either increases or decreases with increasing $Q$, depending on the radius of the emitter. The shift rapidity increases as the electric charge increases when the radius of the emitter is fixed at $r_e=2r_{ISCO}$ for each black hole, whereas increasing $Q$ leads to a decrement in $\dot{z}_{RN}$ if we fix the radius of the emitter for every black hole at $r_e=10M$. In addition, as one can see from this figure, the redshift rapidity at the line of sight is maximal, hence it is easier to be measured at $\varphi=0$.

As the final step, we substitute $r_e$ from the approximation (\ref{rdelta}) in the redshift rapidity (\ref{RRgeneral}) and perform the approximation $\delta \rightarrow 0$ to obtain
\begin{align}\label{zdotmid}
    \dot{z}_{RN}\approx&\frac{m-q^2}{D\sqrt{1-3m+2q^2}(1-2m+q^2)^{3/2}}\times\notag\\
    &\Biggl\{1-\frac{1}{6(1-3m+2q^2)(m-q^2)(1-2m+q^2)}\times\notag\\
    &\Bigl[(-25q^4+52q^2+11)m+(27q^2-28)m^2 \notag\\
    &-12m^3-13q^2-20q^4+8q^6\Bigr]\delta^2\Biggr\},
\end{align}
where $m=M/(D\delta)$, $q=Q/(D\delta)$, and we fixed $\varphi = \pm \pi /2$ in order to find the redshift rapidity of the orbiting particles on the midline.

Now, we have three equations (\ref{m})-(\ref{q}) and (\ref{zdotmid}) for three unknowns $M$, $Q$, and $D$. The rest of the parameters, namely, the total redshift $z_{RN_1}$, the total blueshift $z_{RN_2}$, the redshift rapidity $\dot{z}_{RN}$, and the angular distance $\delta$, are directly measurable quantities. Thus, we can solve Eqs. (\ref{m})-(\ref{q}) and (\ref{zdotmid}) to express \{$M,Q,D$\} in terms of observables \{$z_{RN_{1,2}}, \dot{z}_{RN}, \delta$\}, as we shall show in the coming section.

\section{Disentangle mass, charge, and distance in Reissner-Nordstr\"{o}m
spacetime}
\label{Disentangle}

In this section, we disentangle the mass-to-distance ratio (\ref{m}) and charge-to-distance ratio (\ref{q}), and derive the final expressions for $M$, $Q$, and $D$ in terms of the total frequency shifts, redshift rapidity of photons, and the aperture angle of the telescope. To do so, we substitute (\ref{m}) and (\ref{q}) in Eq. (\ref{zdotmid}), keeping only the terms up to second order in $\delta$, which leads to
\begin{equation}\label{distance}
    D\approx\frac{\sqrt{RB}(R-B)^2}{2(R+B)\dot{z}_{RN}}\biggl[1-\frac{(R+B)^2}{8}\delta^2\biggr],
\end{equation}
which is the distance to the RN black hole. Now, one can replace the distance relation (\ref{distance}) in Eq. (\ref{rdelta}) to obtain the radius of emitter as follows
\begin{equation}\label{approxre}
    r_e\approx\frac{\sqrt{RB}(R-B)^2}{2(R+B)\dot{z}_{RN}}\delta.
\end{equation}

Finally, we substitute (\ref{distance}) in Eqs. (\ref{m}) and (\ref{q}) to obtain the total mass and electric charge of the RN black hole as below

\begin{equation}\label{mass}
    M\approx\frac{\sqrt{RB}(R-B)^2}{2(R+B)\dot{z}_{RN}}\left(1-\frac{2}{RB}+\frac{4}{(R+B)^2}\right)\delta,
\end{equation}
\begin{equation}\label{charge}
    Q\approx\frac{\sqrt{RB}(R-B)^2}{2(R+B)\dot{z}_{RN}}\sqrt{1-\frac{3}{RB}+\frac{8}{(R+B)^2}}\delta,
\end{equation}
just versus observational quantities. Therefore, from Eqs. (\ref{distance}) and (\ref{mass})-(\ref{charge}), we see that the distance to the RN black hole $D$, the total mass $M$, and the electric charge $Q$ of the RN black hole are expressed in terms of directly measurable elements, namely, the total redshift $z_{RN_1}$, the total blueshift $z_{RN_2}$, the redshift rapidity $\dot{z}_{RN}$, and the aperture angle of the telescope $\delta$.

\section{Discussion and final remarks}
\label{COnclusion}

In this work, we have presented analytic relations for the  mass-to-distance ratio and charge-to-distance ratio of the RN black hole in terms of the total frequency shifts on the midline and the aperture angle of the telescope. Then, we computed the derivative of the frequency shift in the RN spacetime with respect to the proper time, called redshift rapidity throughout the text, which is also a general relativistic invariant observable. Next, we employed the mass-to-distance ratio and charge-to-distance ratio formulas as well as the redshift rapidity to express the RN black hole total mass, electric charge, and its distance from the Earth only in terms of observational quantities.

Hence, we have presented concise and elegant analytic formulas for the mass, charge, and distance to the RN black hole in terms of a few directly measurable elements, such as the total redshift, the total blueshift, the redshift rapidity, and the aperture angle of the telescope. These analytic formulas are valid on the midline when the observer is located far away from the source. The results presented in this paper provide new tools to estimate black hole parameters such as the electric charge and set the basis for further developments, like the inclusion of rotation to the black hole and adding the contribution of the cosmological constant to the total frequency shift due to the recession velocity of the host galaxy.

As the final remark, we would like to mention that, in this study, our main goal was to obtain analytic formulas for the black hole mass and charge, and its distance from the Earth from a theoretical point of view. On the other hand, the general formulas of frequency shift and redshift rapidity can be used to estimate these parameters by employing observations from H$_2$O megamaser systems on the accretion disks of supermassive black holes hosted at the core of AGNs in a similar manner as done for 17 supermassive Schwarzschild black holes \cite{towards,Villalobos,deby,Villalobos2}. This is because, in such astrophysical systems, the emitter radius extends in the sub-parsec region from 0.04pc to 0.5pc from the center of the disk, hence located far from the black hole which allows this simple modeling to be applicable and robust. But in the case of the emitters located close enough to the black hole, one has  to face additional difficulties, like the obfuscated determination of positions of photon sources due to the opacity size of the disk, and to incorporate the extreme conditions near the black hole, such as the consequences of the accretion process and the other properties of the disk, both in the theoretical description of the modeling as well as parameter estimation studies.

\section*{Acknowledgments}
All authors are grateful to CONACYT for support under Grant No.
CF-MG-2558591; M.M. also acknowledges SNI and was supported by CONACYT through the postdoctoral Grants No.
31155 and No. 1242413. A.H.-A. thanks SNI and was supported by a VIEP-BUAP grant.

%%%%%%%%%%%%%%%%%%%%%%%%%%%%%%%%%%%%%%%%%%%%%%%%%%%%%%%%%%%%%%%%%%%%%%%%%%%%%%%%%

\end{document}